\documentclass[aps,prd,preprint,superscriptaddress,showpacs,showkeys]{revtex4}
\usepackage{amssymb,amsmath}
\usepackage{graphicx} 
\usepackage{dcolumn}  
\usepackage{epsfig}   
\usepackage{pstricks}
\usepackage{ifpdf}
\usepackage[T1]{fontenc}
\newcommand{\Tr}{ {\mathrm{Tr}\, }}

\begin{document}


\title{Effective locality and chiral symmetry breaking in $QCD$}


%
%
\author{T. Grandou}
\affiliation{Universit\'{e} C\^ote d'Azur,\\ INPHYNI UMR CNRS 7010; 1361 routes des Lucioles, 06560 Valbonne, France}
\email[]{Thierry.Grandou@inln.cnrs.fr}

\author{P.H. Tsang}
\affiliation{Physics Department, Brown University, Providence, RI 02912, USA}
\email[]{peter_tsang@brown.edu}


\date{\today}

\begin{abstract} A few years ago the use of standard functional manipulations was demonstrated to imply an unexpected property satisfied by the fermionic Green's functions of QCD, and called \textit{effective locality}. This feature of QCD is non-perturbative as it results from a full integration of the gluonic degrees of freedom. In this paper, at eikonal and quenching approximation at least, the relation of effective locality to dynamical chiral symmetry breaking is examined\end{abstract}


\maketitle

\section{\label{SEC:1}Introduction}
The two major issues of QCD in its non-perturbative regime are confinement and chiral symmetry breaking. Some years ago a new non-perturbative property of QCD was disclosed and shown to be exact \cite{QCD1,QCD-II}. It is a non-perturbative property in the sense that gluonic degrees of freedom are integrated out from the onset. Therefore, this new property, called effective locality, bears on fermionic Green's functions and it can be summarised as follows.
 \par
 \textit{For any fermionic $2n$-point Green's functions and related amplitudes, the
full gauge-fixed sum of cubic and quartic gluonic interactions,
fermionic loops included, results in a local contact-type interaction. This local interaction is
mediated by a tensorial field which is antisymmetric both in Lorentz and
color indices. Moreover, the resulting sum appears to be fully gauge-fixing independent, that is, gauge-invariant.}
\par
Years ago, a similar property was discovered in an euclidean formulation of the pure Yang-Mills theory where it complied with a form of \emph{duality}, established at least at first non-trivial orders of a semi-classical expansion \cite{Hug}.
\par
Recently, it has been suggested that effective locality could represent the very way non-abelian gauge invariance is realised in the non-perturbative regime of $QCD$, avoiding the intractable and never ending issue of Gribov's copies \cite{Thess}. If this property is really relevant to $QCD$, it should also shed some light on the two fundamental issues just alluded to above, confinement and dynamical chiral symmetry breaking. It is the latter which is the purpose of the current letter.
\par\medskip
Calculations being quite involved, the chiral condensate will be estimated using the quenching approximation and some \emph{light} form of eikonal approximation, admittedly well suited to high energy scattering processes. Present calculations will be carried on for the two-point Greens function, even though it cannot be conclusive in itself, due to a trivial algebraic statement. This obstruction which, besides its algebraic triviality, presents also meaningful aspects, is circumvented by a more involved four-point Green's function calculation whose full details will be given elsewhere \cite{tgpt}.
\par\medskip
The paper is organised as follows. Calculations are set up in the next section, revisiting somewhat the effective locality context where they are carried out. The chiral condensate is defined as the limit $x=y$ of the Green's function $<\bar{\Psi}(x)\Psi(y)>$, and is evaluated in the case of one quark flavour. It is shown that the result is trivialised algebraically. As in the case of massive two-dimensional $QED$, but for different reasons (contrarily to the case of $QED$ where it can be invoked \cite{FGHF}, the \emph{cluster decomposition property} doesn't hold in $QCD$), this difficulty is circumvented by a four-point Green's function calculation to which an element will be borrowed.
\par
The analytically continued Random Matrix treatment of the matter involves a Vandermonde determinant comprising as much as $2^{120}$ monomials, alternate in sign, to be integrated upon. Even to most powerful computers, this seems to be an impossible task to carry out. 
\par
Now, relying on the properties of random matrice's spectra, it is shown how to use \emph{Wigner's semicircle law} so as to frame the overall contribution of so many monomials and to offer an estimate of the chiral condensate.
\par
These results are commented in a conclusion while some extra technical details are deferred to an Appendix in order to alleviate an already technical enough paper.

\section{\label{SEC:2}Calculations}
\subsection{Two terms, one of them vanishing in the chiral limit} 
 At quenching, the relevant expression to be coped with reads,
  \begin{equation}\label{6}
<\bar{\Psi}(x)\Psi (y)>\,= \mathcal{N}  \int d[\chi] \, e^{\frac{i}{4}\int \chi^2} \, \left. e^{\mathfrak{D}_A^{(o)}} \, e^{ + \frac{i}{2}\int{\chi\cdot {F}} +\frac{i}{2}\int{A \cdot \left({D}_{F}^{(0)}\right)^{-1}\!\! \cdot\, A }} \ {G}_{\mathrm{F}}(x_{}, y_{}|A) \right|_{A\rightarrow 0}\,,\end{equation}
where $\mathcal{N}$ stands for a normalisation factor, taken at zero coupling $g=0$, and where, \begin{equation}\mathfrak{D}^{(0)} _{A} =  - \frac{i}{2} \int\mathrm{d}^4x\int\mathrm{d}^4y\,{\frac{\delta}{\delta A(x)} \cdot  {D}_{\mathrm{F}}^{(0)}(x-y) \cdot \frac{\delta}{\delta A(y)} }\,,\end{equation}in which ${D}_{\mathrm{F}}^{(0)}$ is the usual Feynman propagator, while any other choice would yield the same results \cite{tg2}. The functional $G_F(x,y|A)$ can be written with the help of a standard Schwinger/Fradkin representation \cite{Herb},
 \begin{eqnarray}\label{Fradkin}
&& {G}^{\alpha\beta}_{\mathrm{F}}(x,y|A) = i \int_{0}^{\infty}{ds \ e^{-is m^{2}}} \, e^{- \frac{1}{2} \Tr{\ln{\left( 2h \right)}} } \, \int{d[u]} \, e^{ \frac{i}{4} \int_{0}^{s}{ds' \, [u'(s')]^{2} } } \, \delta^{(4)}(x - y + u(s))\nonumber \\  & & \quad \times {\left[ m + i g \gamma^{\alpha\beta}_{\mu} A^{\mu}_{a}(y-u(s)) \lambda^{a} \right]} \, \left( e^{ -ig \int_{0}^{s}{ds' \, u'_{\mu}(s') \, A^{\mu}_{a}(y-u(s')) \, \lambda^{a}} + g \int_{0}^{s}{ds' \sigma^{\mu \nu} \, {F}_{\mu \nu}^{a}(y-u(s')) \, \lambda^{a}}} \right)_{+},
\end{eqnarray}

\noindent where $h(s_{1},s_{2})=\int_{0}^{s}{ds' \, \Theta(s_{1} - s') \Theta(s_{2} - s')}$ and the subscript $+$ indicates Schwinger's proper-time ordering. Ignoring the term involving the $\sigma^{\mu\nu}$ part in view of the eikonal approximation, allows one to deal with the simplified expression of,
 \begin{eqnarray}\label{}
 {G}^{\alpha\beta}_{\mathrm{F}}(x,y|A) &=&  i \int_{0}^{\infty}{ds \ e^{-is m^{2}}} \, e^{- \frac{1}{2} \Tr{\ln{\left( 2h \right)}} } \, \int{d[u]} \, e^{ \frac{i}{4} \int_{0}^{s}{ds' \, [u'(s')]^{2} } } \, \delta^{(4)}(x - y + u(s)) \\ \nonumber & & \quad \times {\left[ m + i g \gamma^{\alpha\beta}_{\mu} A^{\mu}_{a}(x) \lambda^{a} \right]} \, \left( e^{ -ig \int_{0}^{s}{ds' \, u'_{\mu}(s') \, A^{\mu}_{a}(y-u(s')) \, \lambda^{a}} } \right)_{+}
\end{eqnarray}Inserting this expression into (\ref{6}), it is necessary to take the $A^a_\mu$-dependences outside of the ordered exponential so as to permit the functional operator $e^{\mathfrak{D}_A^{(o)}}$ of (\ref{6}) to act upon linear and quadratic $A_\mu$-field dependences. This can be achieved by introducing two extra functional integrations,
\begin{equation}\label{out}
\left(e^{ig\,\!\int_{-\infty}^{+\infty} {\rm{d}}s\,u'(s)^\mu A^a_\mu(y-u(s))\,\lambda^a}\right)_+\!={N}\!\!\int\! {\rm{d}}[\alpha]\!\int\!{\rm{d}}[\Omega]\, e^{-i\!\int_{-\infty}^{+\infty} {\rm{d}}s\,\,\Omega^a(s)\bigl[\alpha^a(s)-g{u'}_\mu (s) A_a^\mu(y-u(s))\bigr]}(e^{i\int_{-\infty}^{+\infty}{\rm{d}}s\,\alpha^a(s)\lambda^a})_+
\end{equation}
Even within the approximations being used, the net expression for the order parameter is rather cumbersome. Written \textit{in extenso} it reads as,
 \begin{eqnarray}\label{extenso}
<\bar{\Psi}(x)\Psi (y)>= \nonumber  i \mathcal{N}\int_{0}^{\infty}{ds \ e^{-is m^{2}}} \, e^{- \frac{1}{2} \Tr{\ln{\left( 2h \right)}} } \\ \nonumber \int{d[u]} \, e^{ \frac{i}{4} \int_{0}^{s}{ds' \, [u'(s')]^{2} } } \, \delta^{(4)}(x - y + u(s)) \\ \nonumber  \Tr\!\int\! {\rm{d}}[\alpha]\!\int\!{\rm{d}}[\Omega]\, e^{-i\!\int {\rm{d}}s'\,\,\Omega^a(s')\alpha^a(s')} \,(e^{i\int_{-\infty}^{+\infty}{\rm{d}}s\,\alpha^a(s)\lambda^a})_+
\\ \int d[\chi] \, e^{\frac{i}{4}\int \chi^2} \,  e^{\mathfrak{D}_A^{(o)}} \, e^{ + \frac{i}{2}\int{\chi\cdot {F}} +\frac{i}{2}\int{A \cdot \left({D}_{F}^{(0)}\right)^{-1}\!\! \cdot\, A }} \,{\left[ m + i g \gamma_{\mu} A^{\mu}_{a}(x) \lambda^{a} \right]} \,  e^{ -ig \int_{0}^{s}{ds' \, \Omega^a(s')u'_{\mu}(s') \, A^{\mu}_{a}(y-u(s')) } } 
\end{eqnarray}
followed, in the last line, by the prescription of $A^a_\mu\rightarrow 0$. The prescription of trace bears on spinorial and color indices. Out of (\ref{extenso}), two contributions are obtained,
\begin{eqnarray}\label{I}
\nonumber  \mathcal{N}\,im\int_{0}^{\infty}{ds \ e^{-is m^{2}}} \, e^{- \frac{1}{2} \Tr{\ln{\left( 2h \right)}} } \\ \nonumber \int{d[u]} \, e^{ \frac{i}{4} \int_{0}^{s}{ds' \, [u'(s')]^{2} } } \, \delta^{(4)}(x - y + u(s)) \\ \nonumber   \Tr\!\int\! {\rm{d}}[\alpha]\!\int\!{\rm{d}}[\Omega]\, e^{-i\!\int {\rm{d}}s'\,\,\Omega^a(s')\alpha^a(s')} \,(e^{i\int_{-\infty}^{+\infty}{\rm{d}}s\,\alpha^a(s)\lambda^a})_+
\\ \int d[\chi] \, e^{\frac{i}{4}\int \chi^2} \,  \left. e^{\mathfrak{D}_{A}^{(0)}} \, e^{{+ \frac{i}{2} \int{ A ^a_\mu\, K^{\mu\nu}_{ab}\, A^b_\nu} }} \, e^{i\int{Q^a_\mu A^\mu_a } }\right|_{\ A \rightarrow 0 }\,,\end{eqnarray}
and,
 \begin{eqnarray}\label{II}
\nonumber  i\mathcal{N}\,\int_{0}^{\infty}{ds \ e^{-is m^{2}}} \, e^{- \frac{1}{2} \Tr{\ln{\left( 2h \right)}} } \\ \nonumber \int{d[u]} \, e^{ \frac{i}{4} \int_{0}^{s}{ds' \, [u'(s')]^{2} } } \, \delta^{(4)}(x - y + u(s)) \\ \nonumber  \Tr\!\int\! {\rm{d}}[\alpha]\!\int\!{\rm{d}}[\Omega]\, e^{-i\!\int {\rm{d}}s'\,\,\Omega^a(s')\alpha^a(s')} \,(e^{i\int_{-\infty}^{+\infty}{\rm{d}}s\,\alpha^a(s)\lambda^a})_+
\\ \int d[\chi] \, e^{\frac{i}{4}\int \chi^2} \,  \left. e^{\mathfrak{D}_{A}^{(0)}} \, \left[ig\gamma^\mu A_\mu^a(x)\lambda^a\right]
e^{{+ \frac{i}{2} \int{ A ^a_\mu\, K^{\mu\nu}_{ab}\, A^b_\nu} }} \, e^{i\int{Q^a_\mu A^\mu_a } }\right|_{\ A \rightarrow 0 }\,,\end{eqnarray}
where in both expressions (\ref{I}) and (\ref{II}) the following definitions apply,
\begin{equation}\label{QK}
K_{\mu\nu}^{ab}=gf^{abc}\chi_{\mu\nu}^c+\left({{D}_{\mathrm{F}}^{(0)}}^{-1}\right)_{\mu \nu}^{a b}\,, \ \ \ Q^a_\mu =-\partial^\nu\chi^a_{\mu\nu}+gR^a_{\mu}\,, \ \  \ f^{abc}\chi^c_{\mu\nu}=
(f\cdot \chi)^{ab}_{\mu\nu}\,,\end{equation}
with,
 \begin{equation}\label{currents}
 {D_F^{(0)}}_{\mu\nu}=g_{\mu\nu}D_F\,, \ \ -\partial^2D_F=\delta^{(4)}\,,\ \ \ R^a_{\mu}(z) = \int_0^s\mathrm{d}s'\,\Omega^a_{}(s')\,u'_\mu(s')\,\delta^4(z - y + u(s'))\,. \end{equation}
 \par\medskip\noindent
  For the first contribution (\ref{I}) one gets for the last line,
 \begin{eqnarray}\label{EL}
& & \left. e^{-\frac{i}{2} \int{\frac{\delta}{\delta A} \cdot {D}_{\mathrm{F}}^{(0)} \cdot  \frac{\delta}{\delta A} }} \cdot e^{+ \frac{i}{2} \int{A \cdot {K} \cdot A} + i \int{A \cdot {Q} }}  \right|_{A \rightarrow 0} \\ \nonumber &=& e^{-\frac{1}{2} \Tr{\ln{\bigl[-g{D}_{\mathrm{F}}^{(0)}\bigr]}}} \cdot \frac{1}{\sqrt{\det(f\cdot\chi)}}\cdot e^{-\frac{i}{2g} \int\mathrm{d}^4z\ {{Q}(z) \cdot (f\cdot\chi(z))^{-1}\cdot {Q}(z)}}\,.
\end{eqnarray}This is the known effective locality phenomenon of the resulting effective interaction \cite{QCD1, QCD-II} where between currents $Q^a_\mu$(z) a local interaction is mediated by the term $(f\cdot\chi)^{-1}(z)$. In the strong coupling limit, $g\gg1$, relevant to the non-perturbative regime of $QCD$, the $\partial^\nu\chi^a_{\mu\nu}$-piece of the current $Q^a_\mu$ can be neglected in front of $R^a_\mu$, and in view of (\ref{QK}) and (\ref{currents}), the effective interaction term reduces to,
\begin{equation}\label{16}
 e^{-\frac{i}{2} g\int\mathrm{d}^4z\int \mathrm{d}s'\int \mathrm{d}s"\,
\Omega^a(s')u'_\mu(s')\,{ \left[f\cdot\chi(z)\right]^{-1}}\Omega^b(s")u'_\nu(s")\delta^{(4)}(z-y+u(s'))\delta^{(4)}(z-y+u(s"))}\,.
\end{equation}
Theorems in \emph{Wiener functional space} \cite{Lapidus}, where Fradkin's variables are taken to live, can be used to evaluate the product of $\delta^{(4)}$ distributions appearing in (\ref{16}) and one gets \cite{QCD6},
\begin{equation}
 e^{-\frac{i}{2} g\frac{\mu^2}{\pi}\,
\Omega^a(0)u'_\mu(0)\,{ \left[f\cdot\chi(y)\right]^{-1}}\Omega^b(0)u'_\nu(0)\,C(|u'_0(0)|,|u'_3(0)|)}
\end{equation}with $C(|u'_0(0)|,|u'_3(0)|)$ a function to be identified to ${1/ |{u}'_{3}(0)|  |{{u}}'_{0}(0)|}$, out of a $4$-point fermionic Green's function calculation carried out at same eikonal approximation \cite{QCD6,tgpt}. In this case, one has $u(s)=sp$ and thus $C(|u'_0(0)|,|u'_3(0)|)={1/ Ep}$, in some reference frame. The interaction is peaked at the value of $s=0$, \cite{QCD1,QCD6}, and the remaining integrations can be carried through with the help of \emph{Random Matrices}, as will be shown below for the second term (\ref{II}). Being regular, these integrations lead to a result on the order of $m$ which vanishes in the chiral limit, $m\rightarrow 0$. Details are given in the appendix. This is in agreement with a long known result established in QED, that scalar electrons/positrons do not produce chiral symmetry breaking \cite{Schwinger}. The same applies here, where contrarily to (\ref{II}), in (\ref{I}) nothing refers to any spinorial structure and thus to any chiral symmetry possible breaking. 
\par
In other words, contribution (\ref{I}) remains on the order of the \emph{explicit} chiral symmetry breaking corresponding to non zero quark masses, and vanishes in the chiral limit where quarks masses are taken to zero.
\par\bigskip
For the second contribution (\ref{II}), the last line  can be evaluated in two different ways leading to the same result,
\begin{equation}
\left. e^{\mathfrak{D}_{A}^{(0)}} \, \left[ig\gamma^\mu A_\mu^a(x)\lambda^a\right]
e^{{+ \frac{i}{2} \int{ A ^a_\mu\, K^{\mu\nu}_{ab}\, A^b_\nu} }} \, e^{i\int{Q^a_\mu A^\mu_a } }\right|_{\ A \rightarrow 0 }=\left.  g\gamma^\mu\lambda^a\frac{\delta}{\delta Q^\mu_a(x)}e^{\mathfrak{D}_{A}^{(0)}}e^{{+ \frac{i}{2} \int{ A ^a_\mu\, K^{\mu\nu}_{ab}\, A^b_\nu} }} \, e^{i\int{Q^a_\mu A^\mu_a } }\right|_{\ A \rightarrow 0 }
\end{equation}

whose evaluation yields,
\begin{equation}
e^{-\frac{1}{2} \Tr{\ln{[-g{D}_{\mathrm{F}}^{(0)}]}}} \frac{1}{\sqrt{\det(f\cdot\chi)}}e^{-\frac{i}{2} \int\mathrm{d}^4z\ {{Q}(z) \cdot (gf\cdot\chi(z))^{-1}\cdot {Q}(z)}}[-ig\gamma^\mu\lambda^a][(gf\cdot\chi(x))^{-1}Q(x)]^a_\mu
\end{equation}
where the first term is to be part of the normalization constant $\mathcal{N}$. As before, the strong coupling limit $g\gg1$ is invoked to simplify (\ref{II}) to,
 \begin{eqnarray}\label{II2}
\nonumber  \mathcal{N} \int_{0}^{\infty}{\mathrm{d}s \ e^{-is m^{2}}} \, e^{- \frac{1}{2} \Tr{\ln{\left( 2h \right)}} } \\ \nonumber \cdot \int{d[u]} \, e^{ \frac{i}{4} \int_{0}^{s}{ds' \, [u'(s')]^{2} } } \, \delta^{(4)}(x - y + u(s)) \\ \nonumber \cdot\ \Tr  \int\! {\rm{d}}[\alpha]\!\int\!{\rm{d}}[\Omega]\, e^{-i\!\int {\rm{d}}s'\,\,\Omega^a(s')\alpha^a(s')} \,(e^{i\int_{-\infty}^{+\infty}{\rm{d}}s\,\alpha^a(s)\lambda^a})_+\\ \cdot\  g\int d[\chi] \, e^{\frac{i}{4}\int \chi^2} \, \frac{1}{\sqrt{\det(f\cdot\chi)}}e^{-\frac{i}{2} g\int\mathrm{d}^4z\ {{R}(z) \cdot (f\cdot\chi(z))^{-1}\cdot {R}(z)}}[\gamma^\mu\lambda^a][(f\cdot\chi(x))^{-1}R(x)]^a_\mu \end{eqnarray}
This is of the form,
\begin{equation}
\int_{0}^{\infty}\!\!{\rm{d}s}\,(\dots)\int{d[u]}\, \delta^{(4)}(x - y + u(s))(\dots)\int\! {\rm{d}}[\alpha]\!\int\!{\rm{d}}[\Omega]\,(\dots)\int d[\chi](\dots)R(x)]_b^\nu 
\end{equation}
and in view of $R(x)]_b^\nu$ given in (\ref{currents}), involves the product of delta distributions,
\begin{equation}\label{involved}
\int_{0}^{\infty}\!\!{\rm{d}s}\int_{0}^{s}\!\!{\rm{d}s'}\,\delta^{(4)}(x - y + u(s))\delta^{(4)}(x - y + u(s'))=\int_{0}^{\infty}{\rm{d}s}\int_{0}^{s}{\rm{d}s'}\,\delta^{(4)}(x - y + u(s))\delta^{(4)}(u(s')-u(s))
\end{equation}
From Wiener space results again, the evaluation of (\ref{involved}) can be inferred from the $4$-point fermionic Green's function's case \cite{QCD6} to read,
\begin{equation}
\int_{0}^{\infty}\!{\rm{d}s}\,(\dots)\int_{0}^{s}\!{\rm{d}s'}\,\delta^{(4)}(x - y + u(s))\delta^{(4)}(u(s')-u(s))=\frac{\mu^2}{\pi Ep}\int_{0}^{\infty}{\rm{d}s}\,\,(\dots)\delta(s)\,\delta^{(4)}(x - y+u(s))
\end{equation}where $\mu^2$ stands for the mass squared associated to the property of effective locality \cite{QCD1, QCD-II, QCD5, QCD6}. Relying on the functional identities (\ref{Id1}) and (\ref{Id2}), it is straight forward to obtain for (\ref{II2}) the expression,
 \begin{eqnarray}\label{II3}
 \mathcal{N}\frac{g\mu^2}{\pi Ep}\,\delta(x-y)   \nonumber  \cdot\ \Tr\!\int\! {\rm{d}}[\alpha(0)]\!\int\!{\rm{d}}[\Omega(0)]\, e^{-i \delta s\,\Omega^a(0)\alpha^a(0)}\, e^{i\delta s\,\alpha^a(0)\lambda^a} \\ 
 \cdot\!\int \mathrm{d}[\chi] \, \frac{e^{\frac{i}{4}\int \chi^2}}{\sqrt{\det(f\cdot\chi(y))}}\,e^{-\frac{i}{2} g\frac{\mu^2}{\pi Ep}\,
\Omega^a(0)u'_\mu(0)\,{ \left[f\cdot\chi(y)\right]^{-1}}\Omega^b(0)u'_\nu(0)}[\gamma^\mu\lambda^a][(f\cdot\chi(x))^{-1}\Omega(0)u'(0)]^a_\mu \end{eqnarray}

The interaction term being peaked at $s=0$, only the contribution at $z=y$ of $\chi^2(z)$ will play a role, all of the other points giving an infinite continuous product of factors $1$ due, at each point, to a normalisation taken at $g=0$, \cite{QCD1, QCD-II, QCD5, QCD6}. The small increment $\delta s$ is introduced so as to keep the exponent dimensionless, and will play no role otherwise (see comment after (\ref{II4})). For the last line of (\ref{II2}) this gives,
\begin{equation}
\int d[\chi]\  e^{\frac{i}{4}\int \mathrm{d}^{(4)}z\ \chi^2(z)} \rightarrow    \int d[\chi(y)]\ e^{\frac{i}{4} \chi(y)^2\Delta}\end{equation}where $\Delta$, of dimension $M^{-4}$, will disappear from the final result (\ref{condens1}). This dimensionful parameter can be used to turn the original $\chi$-field into the dimensionless field, $\bar{\chi}\equiv{\sqrt{\Delta}}\,\chi$. As the normalisation factor $\mathcal{N}$ entails both $\mathrm{d}\chi$ and $\sqrt{ \det(f\cdot\chi)}$, the way in which this elementary space-time cell extension comes into play is read out of the last line of (\ref{II2}),
\begin{equation}\label{Delta}
{\sqrt{\Delta}}\int \mathrm{d}[\chi] \, \frac{e^{\frac{i}{4} \chi^2}}{\sqrt{\det(f\cdot\chi(y))}}\,e^{-\frac{i}{2} g\frac{\mu^2{\sqrt{\Delta}}}{\pi Ep}\,
\Omega^a(0)u'_\mu(0)\,{ \left[f\cdot\chi(y)\right]^{-1}}\Omega^b(0)u'_\nu(0)}[\gamma^\mu\lambda^a][(f\cdot\chi(x))^{-1}\Omega(0)u'(0)]^a_\mu
\end{equation}where the same symbol of $\chi$ is eventually written instead of $\bar{\chi}$. 
\par\medskip
To proceed with a(n analytically continued \cite{QCD6}) Random Matrix integration of (\ref{Delta}) it is convenient to define $\chi^a_{\mu\nu}\otimes T^a=i\mathbb{M}$ where the $T^as$ are the eight Lie algebra generators of $SU_c(3)$ taken in the \emph{adjoint} representation. Matrices $\mathbb{M}$ turn out to be real symmetric traceless matrices of format $N\times N$, that is $32\times 32$, at $D=4$ space-time dimensions and $N_c=3$ colours, as $N=D(N_c^2-1)$. Matrices $i\mathbb{M}$ are also \emph{skew-symmetric}.
\par
Likewise one can define the 32-component vector $V^i={u'}^\mu\otimes\Omega_a$ and the $32\times 32$ matrices $N^i\equiv\gamma^\mu\otimes\lambda^a$, where the $\lambda^as$ are the eight Lie algebra generators of $SU_c(3)$ taken, this time, in the \emph{fundamental} representation, \textit{i.e.}, the Gell-Mann matrices. Also we define ${\hat{\alpha}}^i=[(1,1,1,1)\otimes {\mathbf{\alpha}}]^i$ and ${\hat{\lambda}}^i=[(1,1,1,1)\otimes {\mathbf{\lambda}}]^i$. As recalled in Appendix C, the linear spaces which are constructed in this way are canonically endowed with a metric structure. A one-to-one relation exists between new and original indices. With $n\equiv N_c^2-1$ and $\mu=0,1,2,3$, one has by construction,
\begin{equation}\label{mapping}
(\mu,a)\rightarrow i=a+\mu n\,,\ \ \ \ \ \ i\rightarrow (\mu=\frac{i-i_{m^{(o)}n}}{n},\,a=i_{m^{(o)}n})\,,\end{equation}where the notation $i_{m^{(o)}n}$ stands for \textit{$i$-modulo-n}. In this way (\ref{II2}) can be re-written in a unified \emph{covariant} way as,
\begin{eqnarray}\label{II3}
 i\mathcal{N}\,\delta(x-y)\ \Tr\!\int\! {\rm{d}}[\hat{\alpha}]\,e^{-i\frac{\delta s}{2}\,\hat{\alpha}\cdot\hat{\lambda} } \!\int\!{\rm{d}}[V]\, e^{-i \frac{\delta s}{4\Sigma}\,\, \hat{\alpha}\cdot V} \nonumber\\ 
 \cdot\  N^i\frac{\delta}{\delta V^i}\int \mathrm{d}\mathbb{M} \, \frac{e^{\frac{i}{8N_c}\Tr \mathbb{M}^2(y)}}{\sqrt{\det(\mathbb{M}(y))}}\,e^{-\frac{i}{2} g\frac{\mu^2\sqrt{\Delta}}{\pi Ep}\,
V\cdot \mathbb{M}^{-1}(y)\,\cdot V}\,, \end{eqnarray}where, as will be commented later, the dimensional quantity $\Sigma$ is identified out of a $4$-point calculation \cite{QCD6}. The expression (\ref{II3}) represents the contributions of (\ref{I}) and (\ref{II}) to the chiral limit of $<\bar{\Psi}(x)\Psi (x)>$ at strong coupling, quenching and a \emph{mild} eikonal approximations.  The $mass^{32}$ and $mass^{-64}$ of the integration measures ${\rm{d}}[V]$ and ${\rm{d}}[\hat{\alpha}]$ respectively, simplify to unity with the normalisation $\mathcal{N}$ and leave an order parameter of dimension $(mass)^{3}$ as it should. 
\par\medskip
Now, the matter is that one has,
\begin{equation}\label{trivial}
 \Tr  \,\,e^{-{i\over 2}\,\delta s\,{\hat{\alpha}}\cdot {\hat{\lambda}}}\ N^j=0\ , \ \ \ \forall j=1,\dots, N\,,
\end{equation}where the trace-prescription bears now on $i,j$-indices, $1\leq i,j,\leq 32$, a mix of colour ($a$), and spin indices ($\gamma_\mu^{\alpha\beta}$) in view of the definition of matrices $N^i$. This algebraic statement completely trivialises the order parameter of the chiral symmetry, and it is worth observing that the situation is not improved by relaxing this part of the eikonal approximation. Including the $\sigma_{\nu\mu}$-term of (\ref{Fradkin}), in effect, amounts to redefine vectors $V$s according to,
\begin{equation}
V^i\longrightarrow\tilde{V}^i=(u'_\mu I\!\!\!I_{D\times D}+2i\sigma_{\nu\mu}\partial^\nu_z)\otimes \Omega^a\,,
\end{equation}whose components, with $\sigma_{\nu\mu}=\frac{i}{2}[\gamma_\nu,\gamma_\mu]$, are now matrix-valuated. This additional structure, however, does not improve on the trivial result (\ref{trivial}) since one obtains again,
\begin{equation}\label{ttrivial}
 \Tr  \,\,e^{-i \frac{\delta s}{4}(\frac{E-p}{Ep})\,\, \hat{\alpha}\cdot \tilde{V}} e^{-{i\over 2}\,\delta s\,{\hat{\alpha}}\cdot {\hat{\lambda}}}\ N^j=0\ , \ \ \ \forall j=1,\dots, N\,,
\end{equation}with, again, the factor $(E-p)/Ep$ a $4$-point calculation element \cite{QCD6}. Of course, this does not mean that the order parameter be trivial, but imposes to circumvent the trivial algebraic vanishing of (\ref{trivial}). 

As in massive $QED$ at $D=2$ space-time dimensions \cite{FGHF}, a trick allows to circumvent this issue and consists in calculating the $4$-point fermionic Green's function $<\bar{\Psi}^\alpha(x_1)\Psi _\alpha(y_1)\bar{\Psi}^\beta(x_2)\Psi_\beta (y_2)>$ in the limits of $x_i=y_i$, $i=1,2$. In this way in effect, the trivial left hand side of (\ref{trivial}) gets basically replaced by expressions like $ \Tr (N^j\hat{\lambda}_{j'})(N^k\hat{\lambda}_{k'})$ whose non-triviality can be checked at $j=k$, and $j'=k'$ and/or $j=j'=k'=k$.
\par\medskip
Though straight forward, calculations are quite long and cumbersome and will be presented elsewhere \textit{in extenso} as they appear to testify to non-perturbative $QCD$ peculiar features \cite{tgpt}. As discussed in Section III, the necessity of proceeding through a $4$-point fermionic Green's function calculation in order to obtain the chiral condensate, will be put forth as a possibly meaningful aspect of the non-perturbative regime of $QCD$.
\par
Like in massive two-dimensional $QED$, but for different reasons \cite{FGHF, tgpt}, the $4$-point fermionic Green's function calculation displays the squared value of the chiral condensate. In this letter, accordingly, we will content ourselves of a calculation based on (\ref{II3}), and ignore that it is trivialised by the trace prescription. The difference will come out as some missing accurate $n, N_c$ dependences, to be corrected later \cite{tgpt}, and which fall beyond the scope of our present focus, \textit{i.e.}, the relation of the enigmatic effective locality mass scale, $\mu$, to the dynamical breaking of chiral symmetry. 
\par
Moreover, in the context of these calculations and beyond, that is for the sake of further experimental predictions \cite{PeterThesis},
this will allow one to introduce a manipulation of {Wigner's semicircle law} \cite{Mehta}. \subsection{Using Wigner's semicircle distribution}
In the second line of (\ref{II3}), one has for the measure of integration \cite{Mehta},
\begin{eqnarray}\label{Measure}
& &-i\, {\rm{d}}(\sum_{a=1}^n{\chi^a}_{\mu \nu}\otimes T^a) \nonumber \\ &\equiv& {\rm{d}}\mathbb{M}= {\rm{d}}M_{11}\,{\rm{d}}M_{12} \cdots {\rm{d}}M_{NN} \nonumber \\ &=&\nonumber \left|\frac{ \partial(M_{11}, \cdots, M_{N\!N})}{\partial(\xi_1, \cdots, \xi_N, p_1, \cdots, p_{N(N-1)/2})}\right| \, {\rm{d}}\xi_1 \cdots {\rm{d}}\xi_N \, {\rm{d}}p_1 \cdots {\rm{d}}p_{N(N-1)/2} \\  &=& \prod_{i=1}^{N}\ {\rm{d}}\xi_i  \prod_{i<j} |\xi_i-\xi_j|^{\kappa}\   {\rm{d}}p_1\  ..\ {\rm{d}}p_{N(N-1)/2}\, f(p).
\end{eqnarray}where the $\xi_i$s are the eigenvalues of $\mathbb{M}$ and where $\kappa=1$. In term of this parametrisation of $\mathbb{M}$, (\ref{II3}) can be expressed as,
\begin{eqnarray}\label{II4}
 i\mathcal{N}\,\delta(x-y)\ \Tr\!\int\! {\rm{d}}[\bar{\alpha}]\,e^{-\frac{i}{2}\,\bar{\alpha}\cdot\hat{\lambda} } \!\int\!{\rm{d}}[\bar{V}]\, e^{- \frac{i}{4}\, \bar{\alpha}\cdot \bar{V}} \nonumber\\ 
 \cdot\  \frac{1}{\Sigma}N^i\frac{\delta}{\delta \bar{V}^i}\int_{O_N(\mathbb{R})} \mathrm{d}\mathcal{O}(p)\int_{-\infty}^{+\infty}\!\mathrm{d}\lambda\, \biggl[\prod_1^N \,\int_{-\infty}^{+\infty}\frac{\mathrm{d}\xi_k}{\sqrt{\xi_k}}\,e^{\frac{i}{8N_c}\,\xi_k^2}\biggr]\,\biggl[\prod_{ i<j}^N|\xi_i-\xi_j|^\kappa\biggr] \,e^{i\lambda\left(\sum_1^N\xi_i\right)}\nonumber\\ 
 \cdot\ \exp{\,-i\, \frac{g\mu^2\Sigma^2\sqrt{\Delta}}{2\pi Ep}\,
\,\,{}^t(\mathcal{O}\bar{V})\cdot \mathrm{Diag}(\dots,\frac{1}{\xi_k},\dots)\cdot \mathcal{O}\bar{V}}\,, \end{eqnarray}where $\mathrm{d}\mathcal{O}(p)$, a shorthand for the last factors of (\ref{Measure}), is a \emph{Haar measure} of integration on the orthogonal group $O_N(\mathbb{R})$, \cite{tg}. In (\ref{II4}) the dimensions of $V^i$s and of $\hat{\alpha}^i$s have been taken out from the numerator and denominator $\mathcal{N}$, by defining the dimensionless $N$-vectors $\bar{\alpha}\equiv\delta s \hat{\alpha}$ and $ \bar{V}\equiv V\Sigma^{-1}$, while a subsidiary integration on $\lambda$ is introduced \emph{both} in (\ref{II4}) and in the normalisation factor $\mathcal{N}$, in order to implement the traceless property of $\mathbb{M}$-matrices.
\par
Differentiating with respect to $\bar{V}^i$ and integrating on the orthogonal group $O_N(\mathbb{R})$ \cite{tg}, one obtains,
\begin{eqnarray}\label{II5}
 \mathcal{N}\,\delta(x-y)\,\frac{g\mu^2\Sigma\sqrt{\Delta}}{\pi N Ep}\  \Tr\!\int\! {\rm{d}}[\bar{\alpha}]\,e^{-\frac{i}{2}\,\bar{\alpha}\cdot\hat{\lambda} } \!\int\!{\rm{d}}[\bar{V}]\, e^{- \frac{i}{4}\, \bar{\alpha}\cdot \bar{V}} \nonumber\\ 
 \cdot\  N^i\int_{-\infty}^{+\infty}\!\mathrm{d}\lambda\, \biggl[\prod_1^N \,\int_{-\infty}^{+\infty}\frac{\mathrm{d}\xi_k}{\sqrt{\xi_k}}\,e^{\frac{i}{8N_c}\,\xi_k^2}\biggr]\,\biggl[\prod_{ i<j}^N|\xi_i-\xi_j|^\kappa\biggr]\,\frac{\bar{V}^i}{\xi_i} \,e^{i\lambda\,\sum_1^N\xi_i} 
 \cdot\, \,\left(1+\dots\right)\,, \end{eqnarray}retaining the first non trivial contribution to the expansion of the interaction term in (\ref{II4}). Integrations on $\bar{V}^j$ and then on $\bar{\alpha}^k$ are easily carried out and give (up to the dots),
 \begin{eqnarray}\label{II6}
 -\mathcal{N}\,\delta(x-y)\,\frac{g\mu^2\Sigma\sqrt{\Delta}}{2\pi N Ep}\nonumber\\ 
 \cdot\ \Tr\,  \int_{-\infty}^{+\infty}\!\mathrm{d}\lambda\, \biggl[\prod_1^N \,\int_{-\infty}^{+\infty}\frac{\mathrm{d}\xi_k}{\sqrt{\xi_k}}\,e^{\frac{i}{8N_c}\,\xi_k^2}\biggr]\,\biggl[\prod_{ l<m}^N|\xi_l-\xi_m|^\kappa\biggr]\,\biggl[\sum_i^N\frac{N^i\hat{\lambda}^ig_{ii} }{\xi_i}\biggr] \ e^{i\lambda\,\sum_1^N\xi_i} 
 \,. \end{eqnarray}As shown in Appendix C, the presence of the metric tensor element $g_{ii}$ is crucial to allow the replacement of $\Tr \sum_iN^i\hat{\lambda}^ig_{ii}$ by the constant $-i(N_c-1)$,

  and to write for (\ref{II6}),
  \begin{eqnarray}\label{II7}
 -\mathcal{N}\,\delta(x-y)\,\frac{g\mu^2\Sigma\sqrt{\Delta}}{2\pi Ep}\,({N_c-1})(8N_c)^{\frac{2N^2-N-4}{4}}\nonumber\\ 
 \cdot \int_{-\infty}^{+\infty}\!\mathrm{d}\lambda\   \int_{-\infty}^{+\infty}\frac{\mathrm{d}\xi_1}{\xi_1\sqrt{\xi_1}}\int_{-\infty}^{+\infty}\frac{\mathrm{d}\xi_2}{\sqrt{\xi_2}}\dots\int_{-\infty}^{+\infty}\frac{\mathrm{d}\xi_N}{\sqrt{\xi_N}}\,\prod_{ l<m}^N|\xi_l-\xi_m|^\kappa\, \,e^{{i}\sum_1^N\xi_i^2}\,\cdot\,e^{i\lambda\,\sum_1^N\xi_i}
\end{eqnarray}This result however is vanishing as can be seen in various ways \cite{tgpt}, the simplest being to rely on the structure of $Sp\, \mathbb{M}$, the spectrum of the matrices $\mathbb{M}$ defined in (27), displaying $N/2$ pairs of equal and opposite real eigenvalues \cite{tg}. The expression (\ref{II7}) can accordingly be re-written as \cite{tg},

 \begin{equation}
(\ref{II7})= Constant\cdot \int_{-\infty}^{+\infty}\frac{\mathrm{d}\xi_1}{\xi_1}\int_{-\infty}^{+\infty}{\mathrm{d}\xi_2}\dots\int_{-\infty}^{+\infty}{\mathrm{d}\xi_{N\over 2}}\,\prod_{ l<m}^{N\over 2}(\xi^2_l-\xi^2_m)^2\, \,e^{\,2{i}\sum_1^{N\over 2}\xi_i^2}=0\,,
\end{equation}whose vanishing is manifest, but makes sense only if the same measure of integration on $\mathbb{M}$ does not lead to a vanishing result for the normalisation constant either. This latter condition allowing one to introduce some basic elements of a Random Matrix treatment to be used at next order, $\mathcal{O}(X^2)$, with $X$ defined in Equation (\ref{X.}) below, is analysed in some details now.
\par
 That the normalisation be non-zero can be seen as follows,
 \begin{equation}\label{Z}
\mathcal{N}=\frac{2^N\pi^{N(N+1)\over 4}}{\prod_1^N\Gamma({k\over 2})} \int_{-\infty}^{+\infty}{\mathrm{d}\xi_1}\,\int_{-\infty}^{+\infty}{\mathrm{d}\xi_2}\dots\int_{-\infty}^{+\infty}{\mathrm{d}\xi_{N\over 2}}\,\prod_{ l<m}^{N\over 2}(\xi^2_l-\xi^2_m)^2\, \,e^{\,2{i}\sum_1^{N\over 2}\xi_i^2}\,,
\end{equation}where the prefactor stands for $vol(O_N(\mathbb{R})$, the \emph{volume} of the orthogonal group ${O_N(\mathbb{R})}$ \cite{vol}. After rescaling and analytical continuation \cite{QCD6}, $\xi_i\,\rightarrow \sqrt{i}\,\Theta_i$, with $\xi_i,\Theta_i\in \mathbb{R}$, one obtains,
\begin{equation}\label{Z1}
\mathcal{N}={\sqrt{2}}^{{3N\over 2}-{N^2\over 4}} {\sqrt{i}}^{{N^2\over 4}}\,vol(O_N(\mathbb{R})
 \cdot \int_{0}^{+\infty}{\mathrm{d}\Theta_1}\,\int_{0}^{+\infty}{\mathrm{d}\Theta_2}\dots\int_{0}^{+\infty}{\mathrm{d}\Theta_{N\over 2}}\,\prod_{ l<m}^{N\over 2}(\Theta^2_l-\Theta^2_m)^2\, \,e^{\,{-}\sum_1^{N\over 2}\Theta_i^2}\,.\end{equation}The exponential suppression of any $\Theta_i$-contribution beyond $\Theta_i= 1$ is used now. First, one has clearly,
 \begin{equation}
 \mathcal{N}\geq {\sqrt{2}}^N {\sqrt{i}}^{{N^2\over 4}}\,vol(O_N(\mathbb{R}) \int_{0}^{+\infty}{\mathrm{d}\Theta_1}\,\int_{0}^{+\infty}{\mathrm{d}\Theta_2}\dots\int_{0}^{+\infty}{\mathrm{d}\Theta_{N\over 2}}\,\prod_1^{N\over 2}\Theta_i^{{N\over 2}-1}\prod_{ l<m}^{N\over 2}(\Theta_l-\Theta_m)^2\, \,e^{\,{-}\sum_1^{N\over 2}\Theta_i^2}\,,\end{equation}because $2\Theta_l\Theta_m\leq (\Theta_l+\Theta_m)^2$. Given the symmetry of the integrand one may write eventually, with $K(N)$ the exponent $(\frac{N}{2})^{\frac{N}{2}}$,
  \begin{equation}\label{Major}
 \mathcal{N}\geq {\sqrt{2}}^N {\sqrt{i}}^{{N^2\over 4}}\,vol(O_N(\mathbb{R}) \int_{0}^{+\infty}{\mathrm{d}\Theta_1}\,\,\Theta_1^{K(N)}
 \int_{0}^{+\infty}{\mathrm{d}\Theta_2}\dots\int_{0}^{+\infty}{\mathrm{d}\Theta_{N\over 2}}\,\prod_{ l<m}^{N\over 2}(\Theta_l-\Theta_m)^2\, \,e^{\,{-}\sum_1^{N\over 2}\Theta_i^2}\,.\end{equation} Introducing now the standard Random Matrix definitions,
  \begin{equation}\label{M1}
P_{N\kappa}(\Theta_1,\dots,\Theta_N)\equiv \ C_{N\kappa}\prod_{ i<j}^N|\Theta_l-\Theta_j|^\kappa \,e^{{-}\sum_1^N\Theta_i^2}\,,\end{equation}
\begin{equation}\label{M2}
C^{-1}_{N\kappa}\equiv(2\pi)^{\frac{N}{2}}\,2^{-{N^2\over 2}}\,
\prod_{j=1}^N\Gamma(1+{j})\,, \ \ \ \ \ \mathrm{at}\  \kappa=2\,.\end{equation}
and,
 \begin{equation}\label{M2}
 \biggl[\,\prod_{j=2}^N\int_{-\infty}^{+\infty}\,{\mathrm{d}\Theta_j}\biggr]  \ P_{N2}(\Theta_1,\dots,\Theta_N)\equiv N^{-1}\,\sigma_N(\Theta_1)\end{equation}in accord to another standard definition of Random Matrix theory \cite{Mehta}, one can write (\ref{Major}),
 \begin{eqnarray}\label{Major2}
  \mathcal{N}\geq  \frac{2}{N}{\sqrt{2}}^N\,vol(O_N(\mathbb{R}) \int_{-\infty}^{+\infty}{\mathrm{d}\Theta}\,\, \Theta^{\,\,{N\over 2}^{N\over 2}}\,\sigma_{{N\over 2}}(\Theta)\nonumber\\ \simeq  \frac{2}{N}{\sqrt{2}}^N\,vol(O_N(\mathbb{R}) \int_{-\sqrt{N}}^{+\sqrt{N}}{\mathrm{d}\Theta}\ \Theta^{\,\,{N\over 2}^{N\over 2}}\,\sqrt{N-\Theta^2} \,>\,0\,.
\,\end{eqnarray}where the last and crucial inequality is obvious. Just before it, the approximate equality means `up to sub-leading terms in the large $N$-limit'. In (\ref{Major2}) in effect, use has been made of Wigner's semicircle law, valid at large $N$-values \cite{Mehta},
  \begin{equation}\label{M3}
\sigma_N(\Theta)\longrightarrow \sqrt{2N-\Theta^2}\,, \ \ \mathrm{for}\  -\sqrt{2N}\leq \Theta\leq +\sqrt{2N}\,,\ \  \sigma_N(\Theta)=0\,,\ \mathrm{otherwise\,,}\end{equation}
and in this way the intractable sum of monomials generated by expanding the Vandermonde determinant of (\ref{M1}) is circumvented, while corrections to the asymptotics of (\ref{M3}) can be calculated in a systematic fashion \cite{A8}. This is the more appropriate as the universality of Wigner's semi-circle law is now recognised to  extend far beyond the realm of its original derivation \cite{Krajewski}.

\par\medskip
At next order ($\mathcal{O}(X^2)$, and omitting the bars, \textit{i.e.}, $\bar{V}\rightarrow V$, one  needs the following $O_N(\mathbb{R})$ averages \cite{tg},
 \begin{equation}\label{averages}
 \biggl\langle\sum_1^N\frac{\bigl[(\mathcal{O}V)_j\bigr]^2}{\xi_j}\sum_1^N\frac{1}{ \xi_k} \mathcal{O}_{ki}\mathcal{O}_{kl}V^l\biggr\rangle_{\!O_N(\mathbb{R})}={3\over N^2}\biggl\lbrace V_i(\sum_1^NV_m^2)\,\sum_1^N\frac{1}{\xi_j^2}+V_i^3\sum_1^N\frac{1}{\xi_j}\sum_1^N\frac{1}{\xi_k}\biggr\rbrace\,,
 \end{equation}which result from (\ref{II4}) as  the derivation ${\delta/\delta {V}^i}$ has been performed. Under integration on $Sp\, \mathbb{M}$ the right hand side of (\ref{averages}) reduces to \cite{tgpt},
\begin{equation}\label{reduit}
{3\over N^2}\biggl\lbrace \left(V_i(\sum_1^NV_m^2)\,+V_i^3\right)\,\sum_1^N\frac{1}{\xi_j^2}\biggr\rbrace\,,\end{equation} and one has therefore to evaluate,
 \begin{eqnarray}\label{X2}
 -i\mathcal{N}\,\delta(x-y)\,{\sqrt{8N_c}}^{{N^2\over 2}-1}\,(\frac{X^2}{\Sigma})\,  \nonumber\\ 
 \cdot\ \Tr N^i\int\! {\rm{d}}[{\alpha}]\,e^{-\frac{i}{2}\,{\alpha}\cdot\hat{\lambda} } \!\int\!{\rm{d}}[{V}]\, e^{- \frac{i}{4}\, {\alpha}\cdot {V}}\biggl[ V_i(\sum_1^NV_m^2)+V_i^3\biggr]\nonumber\\  
  \cdot{3\over N^2}\int_{-\infty}^{+\infty}\!\mathrm{d}\lambda\, \biggl[\prod_1^N \,\int_{-\infty}^{+\infty}\frac{\mathrm{d}\xi_k}{\sqrt{\xi_k}}\,e^{{i}\,\xi_k^2}\biggr]\,\biggl[\prod_{ l<m}^N|\xi_l-\xi_m|^\kappa\biggr]\,\biggl[\sum_1^N\frac{1}{\xi_j^2}\biggr] \ e^{i\lambda\,\sum_1^N\xi_i} 
 \,. \end{eqnarray}having rescaled the eigenvalues according to $\xi_i\rightarrow \xi_i/\sqrt{8N_c}$ as in (\ref{II7}). The dimensionless ratio $X$ has been introduced, 
  \begin{equation}\label{X.}X\equiv\frac{g}{8\pi\sqrt{2 N_c}}\,{\mu^2\Sigma^2\sqrt{\Delta}\over Ep}\,,\end{equation}which, even at large enough coupling constant, $g=10\!-\!20$, remains a small number in the range of the $4$-momenta which are relevant to this non-perturbative case (see Appendix E in \cite{tg}). The second line of (\ref{X2}) yields a factor of
  \begin{equation}\label{2ndline} \frac{2}{3}\, 4^6\,\Tr\,N^i{\hat{\lambda}}^i\,g_{ii}\, {\mathbb{I}}_{3\times 3}-32\,\Tr \,N^i({\hat{\lambda}}^i\,g_{ii})^3\, {\mathbb{I}}_{3\times 3}=
 \Tr\left(\frac{2}{3}\, 4^6\,N^i{\hat{\lambda}}^i-32\,N^i({\hat{\lambda}}^i)^3\right) g_{ii}\,{\mathbb{I}}_{3\times 3}\,,\end{equation}and in this paper, one will calculate the contribution due to the first term in the big parenthesis, of order $N^i\hat{\lambda}^i$, because the second one, of order $N^i\hat({\lambda}^i)^3$, represents only a much smaller contribution. Then, by analytically continuing according to $\xi_i\rightarrow \sqrt{i}\,\Theta_i$ with $\xi_i,\Theta_i\in \mathbb{R}$ as was done to pass from (\ref{Z}) to (\ref{Z1}), one gets,
   \begin{eqnarray}\label{X21}
 i\mathcal{N}\,\delta(x-y)\,{\sqrt{8N_c}}^{{N^2\over 2}-1}\,\frac{2\cdot4^6(N_c-1)}{N}\,\frac{X^2}{\Sigma} \nonumber\\  
  \cdot\int_{-\infty}^{+\infty}\!\mathrm{d}\lambda\,  \,\int_{-\infty}^{+\infty}\frac{\mathrm{d}\Theta_1}{{\Theta^2_1}}\int_{-\infty}^{+\infty}\frac{\mathrm{d}\Theta_2\dots \mathrm{d}\Theta_N}{\sqrt{\Theta_1\Theta_2\dots\Theta_N}}\biggl[\prod_{ l<m}^N|\Theta_l-\Theta_m|^\kappa\biggr]\,e^{{-}\sum_1^N\Theta_i^2}\ e^{i\sqrt{i}\lambda\,\sum_1^N\Theta_i} 
 \,. \end{eqnarray}
 This estimate of order $X^2$ satisfies,
   \begin{eqnarray}\label{X22}
(\ref{X21})\geq i\mathcal{N}\,\delta(x-y)\,{\sqrt{8N_c}}^{{N^2\over 2}-1}\,\frac{2\cdot4^6(N_c-1)}{N}\,\frac{X^2}{\Sigma} \nonumber\\  
  \cdot\int_{-\infty}^{+\infty}\!\mathrm{d}\lambda\  \,e^{i\frac{N\lambda^2}{4}}\,\int_{-\infty}^{+\infty}\frac{\mathrm{d}\Theta_1}{\bigr[{\Theta_1-i\sqrt{i}\,\frac{\lambda}{2}}\bigl]^2}\int_{-\infty}^{+\infty}{\mathrm{d}\Theta_2\dots \mathrm{d}\Theta_N}\biggl[\prod_{ l<m}^N|\Theta_l-\Theta_m|^\kappa\biggr]\,e^{{-}\sum_1^N\Theta_i^2}\,,  \end{eqnarray}
  where, in (\ref{X21}), a change of integration variables has been performed, explicitly, $\forall i,\ \Theta_i\rightarrow \Theta_i-i\sqrt{i}\,\frac{\lambda}{2}$, and where $\kappa=1$.
Using now Random Matrix definitions (\ref{M1}), (\ref{M2}) and (\ref{M3}),
  \begin{eqnarray}\label{X23}
(\ref{X21})\geq i\mathcal{N}\,\delta(x-y)\,{\sqrt{8N_c}}^{{N^2\over 2}-1}\,\frac{2\cdot4^6(N_c-1)}{N}\,\frac{X^2}{\Sigma}\,\frac{1}{NC_{N1}} \nonumber\\ 
 \cdot\,\int_{-\infty}^{+\infty} \mathrm{d}\lambda\,\,e^{\,{iN\lambda^2\over 4}}
\int_{-\sqrt{2N}}^{+\sqrt{2N}}\,\frac{\mathrm{d}\Theta_1}{\bigl[\Theta_1-i\sqrt{i}\frac{\lambda}{2}\bigr]^2}\,\sqrt{2N-\Theta_1^2}\,.
\end{eqnarray}

In (\ref{X23}), the two last integrals can be decomposed into a sum $I_1+I_2$, with,
\begin{equation}\label{I1}
I_1(N)\equiv \frac{1}{-i}\int_{-\infty}^{+\infty} \mathrm{d}\lambda\,\,e^{{iN\lambda^2\over 4}}\,\int_{-\sqrt{2N}}^{+\sqrt{2N}}\,\mathrm{d}\Theta\ \frac{\sqrt{2N-\Theta^2}}{i\Theta^2-{\lambda^2\over 4}}
\end{equation}
\begin{equation}\label{I2}
I_2(N)\equiv i\int_{-\infty}^{+\infty} \mathrm{d}\lambda\,\,{\lambda^2\over 2}\,e^{{iN\lambda^2\over 4}}\,\int_{-\sqrt{2N}}^{+\sqrt{2N}}\,\mathrm{d}\Theta\ \frac{\sqrt{2N-\Theta^2}}{\left(i\Theta^2-{\lambda^2\over 4}\right)^2}
\end{equation}

and while $I_2=0$ as proven in the Appendix, one gets for $I_1$ the finite result,
\begin{equation}\label{I1}
I_1(N)= i\pi\,e^{i\pi\over 4}
\,\int_{-\sqrt{2N}}^{+\sqrt{2N}}\,{\mathrm{d}\Theta\over \Theta}\, {\sqrt{2N-\Theta^2}}\ \,\Phi(\Theta\sqrt{N})\,\equiv\, i\pi\,e^{i\pi\over 4}I(N)
\end{equation}where $\Phi(x)$ is the \emph{probability integral} \cite{GR}. This is obtained by using formulae $3.383\,7.$ and $9.254\,1.$ of Ref.\cite{GR}. The normalisation is evaluated with formula $3.381\, 5.$ of \cite{GR} and the same inequality as used for (\ref{X22}). Up to the same overall multiplicative factor of ${\sqrt{8N_c}}^{{N^2\over 2}-1}$ as the one in (\ref{X22}) with which it cancels out, one obtains,  
   \begin{eqnarray}\label{majorDen}
 \mathcal{N}^{-1} \geq  \frac{vol(O_N(\mathbb{R})}{NC_{N1}}\int_{-\infty}^{+\infty}\!\mathrm{d}\lambda\,\,e^{\,i\frac{N\lambda^2}{4}} \int_{-\infty}^{+\infty}{\mathrm{d}\Theta_1}\,\sigma_N(\Theta_1)
 =\frac{vol(O_N(\mathbb{R})}{NC_{N1}}\,\,\pi\,\,e^{{i\pi\over 4}}\,\sqrt{2\pi N}\,. \end{eqnarray}leading for $\langle\bar{\Psi}(x)\Psi(x)\rangle$ to an estimation, 
\begin{equation}\label{condens} 
    \lim_{y=x}\langle\bar{\Psi}(x)\Psi(y)\rangle\,\simeq -\,g^2\,\cdot\,\frac{4^5(N_c-1)}{\sqrt{\pi^5N^3}}\,\frac{I(N)}{vol(O_N(\mathbb{R})}\,\cdot\, \frac{1}{\Sigma}\,\left({\mu^2\Sigma^2\sqrt{\Delta}\over Ep}\right)^2\,\cdot \lim_{y=x}\delta(x-y)\,,
\end{equation}
\par\medskip
Out of a $4$-point calculation relevant to a 2-by-2 scattering quark process and in the same (mild) eikonal approximation, one can identify $\Sigma=\sqrt{E^2-p^2}$ \cite{QCD6}, where $E$ and $p$ are the quark's energies and momenta in their center of mass system, while the elementary space-time cell probed by effective locality has volume $\Delta\simeq E^{-1}p^{-1}\mu^{-2}$, \cite{QCD-II, QCD6, tgpt, tg}. Now, by the same token, this means also that in the limit $y\rightarrow x$, $y$ cannot approach $x$ beyond this elementary space-time cell extension, that is, one must have essentially $\Delta\,\delta(x-y) \simeq 1$, and the chiral condensate reads eventually as,
\begin{equation}\label{condens1} 
    \lim_{y=x}\,\langle\bar{\Psi}(x)\Psi(y)\rangle\,\simeq -\,g^2\,\mu^3\,\cdot\,\frac{\mu}{\sqrt{Ep}}\,{\sqrt{\frac{E^2-p^2}{Ep}}}^{\,\,3}\,\cdot\,\frac{4^5(N_c-1)}{\sqrt{\pi^5N^3}} \,\frac{I(N)}{vol(O_N(\mathbb{R})}\,.
\end{equation}

\par\medskip
 The absolute value of (\ref{condens1}) represents a lower boundary of $|\langle\bar{\Psi}(x)\Psi(x)\rangle |$. As can be proven easily \cite{tgpt}, an upper and \emph{finite} boundary value can be found also for $|\langle\bar{\Psi}(x)\Psi(x)\rangle|$ and matters to make sure that one deals with a definite quantity. This estimate of the chiral condensate comes out as a product of three groups of terms which can be given the following comments. 
\par
(i) First and most importantly, effective locality involves dynamical chiral symmetry breaking (through non \emph{normal-ordered} Green's function calculations \cite{jamin, sjbetal}).
\par
(ii) The chiral condensate goes like the third power of the effective locality mass scale $\mu$, suggesting a close connection between effective locality and chiral symmetry breaking. 
\par
(iii) Then one observes something one could dub \emph{a partonic depleting function}, here expressed in the center of mass system of a two quark scattering processs, \textit{i.e.}, identified out of a $4$-point Green's function \cite{QCD6, tgpt}. One can see that the chiral condensate magnitude is modulated by a function of energy and momentum variables $E$ and $p$ attached to the quarks, taken as particles (partons). Would the latter variables be in a range of magnitude corresponding to the perturbative regime of QCD, \textit{i.e.}, $E,p\gg\Lambda_{QCD}$ say, then the particle-like character of quarks would allow $\Sigma^2$ to be replaced by $m^2$, the squared quark mass.
\par\noindent
 What (\ref{condens1}) suggests anyway is that as the energy $E$ and momentum $p$ of the scattering quarks increase, the chiral condensate gets substantially depleted in agreement with the fact that chiral symmetry breaking does not result from a perturbative mechanism.
\par
(iv) Eventually comes an involved combination of numerical factors depending on $N_c$ and $D$ ($N=D(N_c^2-1)$). Now, approximations have been used as well as, in the end, Wigner's semi-circle law which applies to very large values of $N$. What is worth pointing out, however, is that using Wigner's law allows one to frame the contribution attached to the intractable sum over the too large number of monomials coming from a Vandermonde determinant expansion, and that corrections to this law can be calculated in a systematic way \cite {A8}. Numerical estimates could be given elsewhere \cite{tgpt}, as the refining of the integration measure on $\mathbb{M}$- matrices will allow better estimates of both (\ref{X21}) and (\ref{majorDen}).

\section{Conclusion}
Remarks are in order concerning the result (\ref{condens1}) and the four comments given above. 
\par\medskip
- Effective locality, an exact non-perturbative property of $QCD$ \cite{QCD-II}, involves dynamical chiral symmetry breaking. In this calculation this is established in the case of one flavour of quark, at large coupling, quenching and a partial eikonal approximation. While the latter is recognised relevant to the matter of high energy scattering, the quenching approximation shouldn't be taken as a proviso to the result. In effect, as supported by models \cite{hftg}, numerical simulations and other non-perturbative approaches such as the Dyakonov instanton calculus \cite{Dyakonov}, whenever chiral symmetry breaking can be established, it is an outcome of \emph{both} quenched and non-quenched calculations.
\par\medskip
- Since it was discovered \cite{QCD1}, $QCD$ effective locality comes about with a mass scale, $\mu$, necessarily associated to it in order to avoid triviality \cite{QCD11}. The relation of non-perturbative phenomena to this mass scale is of utmost interest, and motivated the current analysis. The fermionic condensate, a measure of dynamical chiral symmetry breaking, comes out to be on the order of the third power of it, a reasonable answer in view of a pure dimensional argument. However, estimation (\ref{condens1}) tells more.
\par\medskip
- The result's magnitude is modulated by a decreasing function of $E$ and $p$, the energy and momentum of scattering quarks. The more $E,p$ increase, that is, the more `partonic' the situation becomes, the more the chiral condensate is depleted, down to zero, recovering the perturbative result. This is also in full agreement with experimental facts according to which non-perturbative aspects are washed out beyond an energy scale of about $400\, MeV$~\cite{Kos}.
\par
In the formal context of effective locality, it is remarkable that there is no way to identify such a depleting function out of a $2$-point Green's function calculation, such as $\langle\bar{\Psi}(x)\Psi(y)\rangle$, and this, irrespectively of the `algebraic trivialisation' of the result which is inherent to the $2$-point case and is circumvented by a $4$-point calculation (in the context of effective locality, in effect, a $2$-point calculation only involves a self-interaction whose renormalisation is achieved perturbatively along the lines described in \cite{QCD5'} and in agreement with a prediction of \emph{Light Front $QCD$} \cite{SJB}). 
\par
One can venture the idea that this fact, not innocuous, may be meaningful.

\par\noindent
    Effective locality shows up once gluonic degrees of freedom have been integrated out, and the resulting effective interaction between quarks is then given by \textit{$2$-by-$2$ body} local interactions~\cite{tgD}. This is clearly reminiscent of the reduction of low-energy $QCD$ to \emph{Quark Flavour Dynamics} \cite{HugoR, Schaden}, itself reminiscent of an extended Nambu-Jona Lasinio model \cite{NJL}. Thus, if this is the relevant quark dynamical structure in the non-perturbative regime of $QCD$, it should not come as a surprise that dynamical chiral symmetry breaking be revealed by a $4$-point fermionic Green's function calculation. 
\par\noindent
In some sense, it is as if effective locality did not favour much the long held point of view relating dynamical chiral symmetry breaking to the quark dressed propagator, in particular to its massive pole \cite{sjbetal} (in this respect it is worth observing also that in a most recent axiomatic analysis, the definite existence of such massive poles has been recognised a questionable, gauge-dependent issue \cite{PL}). Quite on the contrary, this very aspect of dynamical chiral symmetry breaking would appear to be consistent with the manner \emph{confinement} seems to show up in this same effective locality context, as could be argued elsewhere \cite{QCD11}.

\par\medskip
- So far, out of effective locality, some qualitative features could be derived from the general form of fermionic $2n$-point Green's functions \cite{QCD-II,QCD6, tg, tg2}, whereas phenomenological applications, however promising, required further simplifications \cite{QCD5,QCD5'}. A major issue has been the very big number of monomials comprised in the Vandermonde determinant (\ref{Measure}), alternate in sign, whose global behaviour when summed up is difficult to appreciate. The present calculation, though a bit technical, avoids this issue by relying on Wigner's semi-circle law. In principle, the latter applies to the cases of very large $N$-values while the current value of $N$ is only $32$. It would be interesting to know how fast, as a function of $N$, does the semi-circle law `saturates', as this law would matter to improve on possible phenomenological applications. Though a bit lengthy, it is to be noted that corrections to the asymptotics of Wigner's law can also be dealt with analytically. This and other related improvements will be coped with in a future publication, when some approximations will be relaxed \cite{tgpt}.
\par
For our present concern, relying on Wigner's law allows one to prove the important fact that $QCD$ dynamical chiral symmetry breaking is an outcome of effective locality, at least within a set of approximations known to preserve chiral symmetry breaking if any; and accordingly, to continue proposing effective locality as a possibly interesting access to the non-perturbative regime of $QCD$.

\section{appendix}
\subsection{On the chiral limit of Equation (\ref{I})} 
In Section II, it is stated that in the chiral limit of $m\rightarrow 0$, no contribution is generated by the term (\ref{I}). This can be seen along the following steps. One can rely on a functional identity,
\begin{equation}\label{Id1}
 e^{ \frac{i}{4} \int_{0}^{s}{ds' \, [u'(s')]^{2} } } =\int{d[u]} \, e^{ \frac{i}{4} \int_{0}^{s}{ds_1}\int_{0}^{s}{ds_2}u(s_1)h^{-1}(s_1,s_2)u(s_2)}\end{equation}where $h^{-1}(s_1,s_2)=\frac{\partial}{\partial s_1}\frac{\partial}{\partial s_2}\delta(s_1-s_2)$, and a representation of the $\delta(x-y+u(s))$ distribution,
 \begin{equation}
 \delta^{(4)}(x-y+u(s))=\int\frac{d^4k}{(2\pi)^4}\, e^{ik\cdot(x-y+u(s))}\,.\end{equation}This allows to take advantage of another functional identity,
 \begin{equation}\label{Id2}
 \int{d[u]} \, e^{ \frac{i}{4} \int_{0}^{s}{ds_1}\int_{0}^{s}{ds_2}u(s_1)h^{-1}(s_1,s_2)u(s_2) +ik\cdot u(s)}=e^{-isk^2}\, e^{+\frac{1}{2}\Tr\ln(2h)}\,,\end{equation}and to write the first contribution (\ref{I}) as,
 \begin{eqnarray}\label{I3}
 im\,\mathcal{N}\int\frac{d^4k}{(2\pi)^4}\, e^{ik\cdot(x-y)}\,\int_{0}^{\infty}ds \ e^{+is\left(k^2-m^2+i\varepsilon\right)}\nonumber
 \\ \nonumber  \Tr\!\int\! {\rm{d}}[\alpha]\!\int\!{\rm{d}}[\Omega]\, e^{-i\!\int {\rm{d}}s'\,\,\Omega^a(s')\alpha^a(s')} \,(e^{i\int_{-\infty}^{+\infty}{\rm{d}}s\,\alpha^a(s)\lambda^a})_+ \\ \int d[\chi] \, e^{\frac{i}{4}\int \chi^2} \,  \frac{1}{\sqrt{\det(f\cdot\chi)}}\,e^{-\frac{i}{2} g\frac{\mu^2}{\pi Ep}\,
\Omega^a(0)u'_\mu(0)\,{ \left[f\cdot\chi(y)\right]^{-1}}\Omega^b(0)u'_\nu(0)}\end{eqnarray}where the Fradkin's field variables $u(s)$ are not treated on the same footing as their derivatives, $u'(s)$, which is consistent with the assumption of taking them to live in a Wiener functional space. Now as argued in the main text, the interaction is peaked at $s=0$ and (\ref{I3}) is reduced to,
 \begin{eqnarray}\label{I4}
 im\,\mathcal{N}\int\frac{d^4k}{(2\pi)^4}\, {e^{ik\cdot(x-y)}\over k^2-m^2+i\varepsilon}\nonumber
 \\ \nonumber  \cdot\ \Tr\!\int\! {\rm{d}}[\alpha(0)]\!\int\!{\rm{d}}[\Omega(0)]\, e^{-i \delta s\,\Omega^a(0)\alpha^a(0)}\, e^{i\delta s\,\alpha^a(0)\lambda^a} \\ \cdot\int d[\chi] \, e^{\frac{i}{4}\int \chi^2} \,  \frac{1}{\sqrt{\det(f\cdot\chi)}}\,e^{-\frac{i}{2} g\frac{\mu^2}{\pi Ep}\,
\Omega^a(0)u'_\mu(0)\,{ \left[f\cdot\chi(y)\right]^{-1}}\Omega^b(0)u'_\nu(0)}\end{eqnarray}As proven in the main text the integration on the $\chi$-fields is regular. In the limit $x=y$, the first line is divergent and is taken care of by renormalisation. Therefore, as the quark mass parameter $m$ is finally taken to zero, the contribution of (\ref{I4}) and thus of (\ref{I}) vanishes.

\subsection{On the vanishing of $I_2$}
Recalling $I_2$,
\begin{equation*}
I_2(N)\equiv \int_{-\infty}^{+\infty} \mathrm{d}\lambda\,\,{\lambda^2\over 2}\,e^{{iN\lambda^2\over 4}}\,\int_{-\sqrt{2N}}^{+\sqrt{2N}}\,\mathrm{d}\Theta\ \frac{\sqrt{2N-\Theta^2}}{(i\Theta^2-{\lambda^2\over 4})^2}\,,
\end{equation*}integration on $\Theta$ can be carried out with the help of formula $3.681\,2.$ in \cite{GR}. One gets,
\begin{equation}
\int_{-\sqrt{2N}}^{+\sqrt{2N}}\,\mathrm{d}\Theta\ \frac{\sqrt{2N-\Theta^2}}{(i\Theta^2-{\lambda^2\over 4})^2}=-{8\pi N\over \lambda^3}\cdot{1\over \sqrt{\lambda^2-8iN}}\end{equation}whose subsequent integration on $\lambda$ is vanishing.

\subsection{On Equation (\ref{II6})}
In the construction outlined right after (\ref{Delta}), which aims also at dealing with the required calculations in  some unified and \emph{covariant} way, the tensorial product space of vectors $V=u'\otimes \Omega$ has dimension $N=32$, and from the metrics on the composing spaces, is endowed with a metric tensor  $g_{ii}\equiv \bigr[(g_{\mu\nu})\otimes (\delta_{ab})\bigl]_{ii}$ which inherits the \emph{Minkowskian} character of spacetime, $g_{\mu\nu}=Diag (1,-1,-1,-1)$. That is, it is a diagonal metric tensor with $8$ components $+1$ and $24$ components $-1$. This is at the origin of a point which would not come about with a pure \emph{Euclidean} metric tensor but requires some care in the current pseudo-euclidean case.

\par
In (\ref{II5}), it is easy to check that the integration,
\begin{equation}
\int_{-\infty}^{+\infty}{\rm{d}}{V}_i\, {V}_i\ e^{- \frac{i}{4}\, {\alpha}_i {V}_i\,g^{ii}}=i{4}^2{g^{ii}}\,\delta'({\alpha}_i)\,, 
\end{equation}does depend on the sign of $g^{ii}$. This dependence, erased within the subsequent integration on ${\alpha}_i$, pops out again in the end, to wit,
\begin{equation}
i{4}^2{g^{ii}}\,\Tr N^i\int_{-\infty}^{+\infty}{\rm{d}}{\alpha}_i\ e^{- \frac{i}{2}\, {\alpha}_i \hat{\lambda}_i\,g^{ii}}\cdot \delta'({\alpha}_i)=-{4^2\over 2}\,(g^{ii})^2\ \Tr \left(N^i\hat{\lambda}_i=N^i\hat{\lambda}^i\,g_{ii}\right)\ {\mathbb{I}}_{3\times 3}\,,
\end{equation}which of course is zero in view of (\ref{trivial}). In the $4$-point calculation designed to circumvent this algebraic triviality, what comes about, as announced after (\ref{ttrivial}), is the trace $ \Tr (\sum_iN^i\hat{\lambda}^{i}\,g_{ii})(\sum_jN^j\hat{\lambda}^{j}\,g_{jj})$ whose only cases of non-triviality require $i=j$. Now, in this way one recovers independence on the sign of $g_{ii}$ since  $g_{ii}^2=+1$ for all $i=1,2,\dots,N$. As a result, once summed upon $i$, this allows to replace the vanishing $ \Tr (\sum_iN^i\hat{\lambda}^{i}\,g_{ii})$ by $-i(N_c-1)$, as is claimed after (\ref{II6}).

 \par

\end{document}